# MULTI-TASK SEMI-SUPERVISED LEARNING FOR PULMONARY LOBE SEGMENTATION

*Jingnan Jia[1], Zhiwei Zhai[1], M. Els Bakker[1], I. Hernández Girón[1], Marius Staring[1], Berend C. Stoel[1*]*

[1] Division of Image Processing, Department of Radiology, Leiden University Medical Center (LUMC), P.O. Box 9600, 2300 RC, Leiden, The Netherlands.

## ABSTRACT

Pulmonary lobe segmentation is an important preprocessing task for the analysis of lung diseases. Traditional methods relying on fissure detection or other anatomical features, such as the distribution of pulmonary vessels and airways, could provide reasonably accurate lobe segmentations. Deep learning based methods can outperform these traditional approaches, but require large datasets. Deep multi-task learning is expected to utilize labels of multiple different structures. However, commonly such labels are distributed over multiple datasets. In this paper, we proposed a multi-task semi-supervised model that can leverage information of multiple structures from unannotated datasets and datasets annotated with different structures. A focused alternating training strategy is presented to balance the different tasks. We evaluated the trained model on an external independent CT dataset. The results show that our model significantly outperforms single-task alternatives, improving the mean surface distance from 7.174 mm to 4.196 mm. We also demonstrated that our approach is successful for different network architectures as backbones.

## 1. INTRODUCTION

Computed Tomography (CT) plays an important role in the diagnosis of lung diseases. The appearance of these diseases is, however, diverse and complex. Some lung diseases predominantly affect certain lobes [1], as each lobe has an independent airway and vascular system. Therefore, lobe segmentation is an important preprocessing step in the automated interpretation of lung CT, in order to quantify lung disease in specific regions of interest.

Traditional lobe segmentation algorithms [2] combine information from fissures, bronchi and pulmonary vessels. Recently, deep neural networks (DNNs) improved the performance in lobe segmentation considerably [3, 4]. However, most of them require hundreds [4] or even thousands [3] of annotated CT scans, which is both laborious and time-consuming. Since the lack of large annotated datasets is a common challenge in medical imaging research, we propose to pool several smaller datasets, each with annotations of different structures, and present a strategy to leverage these multiple annotations.

Different methods utilizing multiple annotations have been proposed, based on multi-task or semi-supervised learning. A multi-task models were proposed to extract multi-label information from one dataset [5], or to pool several different datasets of different organs for pre-training or transfer learning [6]. Semi-supervised learning was proposed for brain MRI segmentation, where a segmentation network and a reconstruction network, sharing the same encoder, were trained with annotated data and unannotated data, respectively [7].

In this paper, we combined multi-task and semi-supervised learning for lobe segmentation. Since the distribution of vessels can help improve lobe segmentation, auxiliary chest CT datasets with vessel annotations were added to train our model. This requires however a proper balance between the training of the different subnets.

The main contributions of this paper are therefore: 1) A multi-task semi-supervised network for lobe segmentation that can utilize information from distinct datasets with annotations of different anatomical structures; and 2) A focused alternating training strategy to let the model train different tasks alternatively on different datasets, and still focus on the main task.

## 2. METHOD

### 2.1. Model design

Our network is composed of three subnets, sharing the same encoder (Fig. 1): two segmentation nets for segmenting lobes and vessels (called LobeNet and VesselNet, respectively) and one reconstruction net (ReconNet). The two segmentation nets are adapted from V-Nets [8], consisting of an encoder and decoder (see Fig. 2). The only difference between the two segmentation nets is the number of output channels of the last layer. LobeNet outputs six channels, representing one background and five lobes. VesselNet has two output channels, representing background and vessels. The ReconNet architecture is similar to the segmentation nets. It omits, however, any skip connections from encoder to decoder, in order to introduce an information bottleneck.

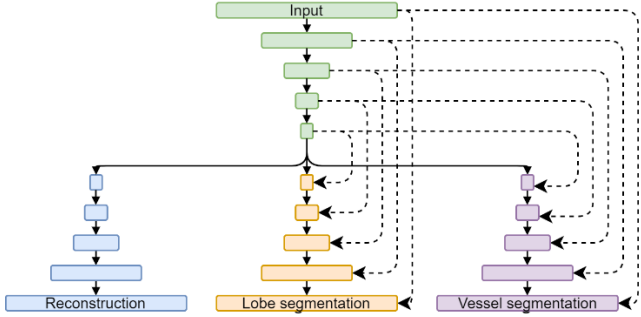

**Fig. 1**. Framework of the proposed model. Solid lines represent convolution paths; dashed lines denote skip connections.

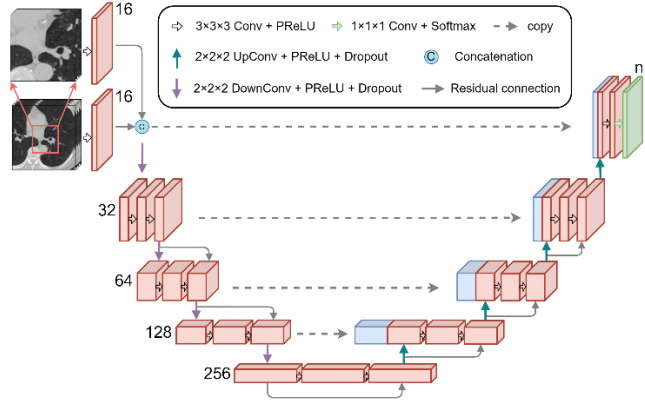

**Fig. 2.** Architecture of the segmentation subnets.

The model is fed with a pair of 3D patches at different fields of view, to provide local and global context (see Fig. 2). One patch is cropped from the CT images with the original resolution, and the other is cropped from a down-sampled image with the same center as the first one.

### 2.2. Loss function and evaluation metrics

Because of the imbalanced labels in the lobe and vessel datasets, the two segmentation tasks were trained using weighted Dices: $loss_{seg} = 1 - \sum_{i=1}^{m} \frac{Dice_i}{V_i}$, where $Dice_i$ represents the Dice of $i^{th}$ class, and $V_i$ represents the volume of $i^{th}$ class. The mean squared error (MSE) was used as the loss function to reconstruct the input images.

To evaluate the model performance, the mean surface distance (MSD), averaged 95$^{th}$ percentile Hausdorff distance (HD95) and average Dice similarity coefficient (DSC) with standard deviation (STD) were calculated by a publicly available tool [9]. To test the significance of the differences between two pairs of results, a Wilcoxon signed rank test was used.

### 2.3 Training strategies

Since the dataset of each subnet has its own annotation, the three subnets cannot be trained jointly. Therefore, we examined two training strategies, where different subnets are trained alternatingly with different datasets (see Fig. 3).

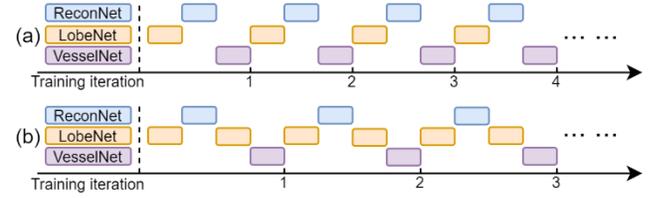

**Fig. 3.** Training strategies. (a) equally alternating training; (b) focused alternating training.

The first training strategy is called *equally alternating training* (EAT), where the gradients of each subnet are updated once during each iteration. In the second training strategy, the gradients of the main subnet (LobeNet) is updated more frequently (twice in each iteration), i.e. after finishing one of the auxiliary subnets (VesselNet or ReconNet). Since this strategy focuses more on the LobeNet, we called this *focused alternating training* (FAT).

### 2.4 Learning rate optimization

Training a multi-task network is challenging, because different tasks have different loss scales and require different learning rates. Balancing different tasks is required so that they can help the main task instead of competing against it. Therefore, we introduce an adaptive learning rate to ensure the step sizes for the auxiliary tasks are always smaller than for the main task.

The learning rate of LobeNet was fixed as 1e-4, whereas for VesselNet and ReconNet they were initialized at 1e-5, and subsequently updated adaptively during training: $lr = \lambda \cdot lr_{lobe} \cdot \frac{loss_{lobe}}{loss}$, where $\lambda$ controls the adaptive rate. In this paper $\lambda = 0.1$ was used.

## 3. DATASET AND DATA AUGMENTATION

### 3.1 Dataset

Four datasets were used: two separate datasets from our institution, **GLUCOLD** [10] and **SSc** [11], were used for segmenting lobes and lung vessels, respectively; one dataset, **LUNA16** [12], for the reconstruction task; and an independent dataset, **LOLA11** [10], was used for external testing.

**GLUCOLD** consists of 22 CT scans of COPD patients. Masks were generated by region growing and further corrected by two experienced researchers. The data was divided into three subsets for training (17), validation (1) and testing (4).

**SSc** contains 77 cases with systemic sclerosis. A Graph-cuts method [13] was applied to obtain initial vessel masks. We selected 55 high quality segmentations to form the dataset for VesselNet, and divided them into training (50) and test data (5). As our main goal was to segment lobes, small errors in vessel masks were considered acceptable.

**LUNA16** is composed of 888 CT scans, selected from the LIDC dataset, initially intended for lung nodule analysis. By excluding all annotations, only the CT data was used for unsupervised learning in ReconNet. Please note that in principle more images could be added into the training dataset for ReconNet since it does not require annotations.

**LOLA11** includes 55 CT scans from various sites. It was not used for training or validation and serves as an independent performance evaluation. LOLA11 organizers annotated visible-only fissures on 9 coronal slices in each case [10]. Our expert (M.E.B) verified the annotations, and corrected them where needed. MSD was calculated based on those slices where fissures were annotated.

### 3.2 Data augmentation

Because of GPU memory limitations, it was not feasible to input the whole high-resolution 3D CT images into the model. Therefore, two 3D patches of size 144×144×96 voxels with different scales were extracted from the CT images. The images of the training dataset were augmented on-the-fly by linear transformations including random shifts (±5%), rotations (±5%), shearing (±5%), and scaling (±5%).

## 4. EXPERIMENTS AND RESULTS

### 4.1 Implementation details

The model was implemented using TensorFlow 1.15. The Adam optimizer was used, batch size was set to 1, and the total number of training steps for main task (LobeNet) was fixed to 100,000. Multithreading was applied to accelerate data preprocessing. The code to replicate the experiments has been released at https://github.com/Ordgod/lobeseg.

Training and validation was performed on an Intel(R) Xeon(R) CPU Gold 6126 @ 2.6GHz machine with 90 GB memory. A single GPU NVIDIA GeForce RTX 2080TI with 11 GB memory was used to accelerate training.

### 4.2 Impact of multi-scale input & adaptive learning rate

We first trained a single LobeNet with single-patch input as a baseline. Then the different techniques were applied to improve it and their effects were examined. The performance of the different techniques is shown in Table 1. It can be seen that compared with single-task learning, all multi-task models achieved significant improvements to varying extents. This confirms that information from other anatomical structures can improve lobe segmentation. Among these multi-task models, the model with adaptive learning rate and multi-scale input performed best.

**Table 1.** Performance (DSC/MSD ± STD) of the different models in GLUCOLD. MI: Multi-scale input. AL: adaptive learning rate. ST: single-task training. MT: multi-task training with FAT. **Bold** indicates the best performance.

| MI | AL | DSC | | MSD (mm) | |
|---|---|---|---|---|---|
| | | ST | MT | ST | MT |
| | | 0.952 ±0.022 | 0.964 ± 0.019 | 1.834 ±0.840 | 1.025 ± 0.480 |
| | √ | | 0.967 ± 0.017 | | 0.868 ± 0.369 |
| √ | | 0.960 ±0.018 | 0.969 ± 0.015 | 1.184 ±0.325 | 1.003 ± 0.366 |
| √ | √ | | **0.970 ± 0.016** | | **0.765 ± 0.294** |

### 4.3 Impact of auxiliary tasks & training strategies

Based on multi-scale input and adaptive learning rate, we progressively introduced different auxiliary tasks. The results are presented in Table 2. It can be observed that the introduction of VesselNet or ReconNet alone can improve the performance of LobeNet to varying extents. The combination of the three subnets performed better than single LobeNet but worse than LobeNet + VesselNet if the equally alternating training strategy was used. The focused alternating training achieved the best performance. Therefore, focusing to the main task was successful, while still making use of the auxiliary tasks.

**Table 2.** Performance of combinations of different tasks with multi-scale input and adaptive learning rate on GLUCOLD. VS: VesselNet. RC: ReconNet.

| Architecture | DSC | MSD (mm) |
|---|---|---|
| LobeNet | 0.960 ± 0.018 | 1.834 ± 0.325 |
| LobeNet +RC | 0.960 ± 0.027 | 1.541 ± 0.513 |
| LobeNet +VS | 0.968 ± 0.015 | 0.827 ± 0.339 |
| LobeNet +RC+VS (EAT) | 0.964 ± 0.017 | 1.032 ± 0.495 |
| LobeNet +RC+VS (FAT) | **0.970 ± 0.016** | **0.765 ± 0.294** |

### 4.4 Comparison with existing networks

To compare our model with existing methods, we replaced LobeNet by FRV-Net [5] and applied multi-task training strategies to it. The models were tested independently on the corrected LOLA11 (Table 3). Compared with the single-task LobeNet, our model achieved a significant ($p<0.005$) improvement. Furthermore, compared with FRV-Net and its multi-task version, our proposed model still achieved a competitive performance. Moreover, our proposed multi-task training strategies also worked for FRV-Net to some extent. This indicates the generalizability of our methods.

**Table 3.** Comparison of single-task and multi-task models with LobeNet or FRV-Net as the main subnet. † denotes significantly better than the single-task model ($p<0.05$).

| Architecture | MSD (mm) | HD95 (mm) | Architecture | MSD (mm) | HD95 (mm) |
|---|---|---|---|---|---|
| LobeNet | 7.124 ±3.432 | 27.980 ±11.908 | FRV-Net | 5.847 ±2.990 | 27.944 ±15.541 |
| LobeNet + RC + VS | **4.696†** ±2.813 | **19.383†** ±9.904 | FRV-Net + RC + VS | 5.135 ±3.705 | 22.528 ±17.269 |

Fig. 4 and Fig. 5 show the qualitative results of different networks. From the 3D and 2D views, it can be seen that the initial segmentation from LobeNet has significant false positive voxels outside the lungs. The segmentation from FRV-Net could not predict the oblique fissure exactly. Our model was able to precisely predict both fissures and lung surfaces.

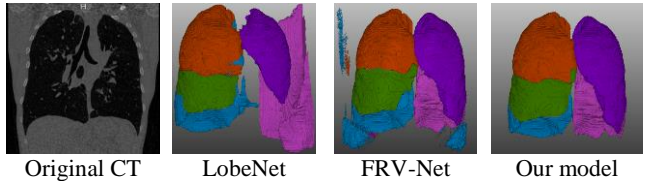

Original CT     LobeNet     FRV-Net     Our model
**Fig. 4.** 3D view of results from one example in LOLA11

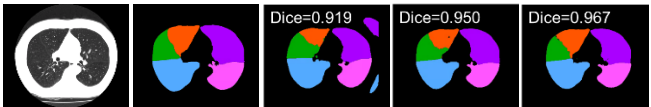

Original CT   Ground truth   LobeNet   FRV-Net   Our model
**Fig. 5.** 2D view of results from one example in GLUCOLD

## 5. CONCLUSIONS

In this paper, a multi-task semi-supervised learning model was proposed for pulmonary lobe segmentation, which can utilize unannotated datasets and annotated datasets of different anatomical structures. Multi-scale input, adaptive learning rate and focused alternating training strategy were introduced to balance different tasks. Experiments show that our multi-task semi-supervised model outperformed single-task LobeNet and FRV-Net. Considering it has been successfully implemented on LobeNet and FRV-Net, our model along with the training methods shows the potential to improve other DNN architectures as well.

## 6. COMPLIANCE WITH ETHICAL STANDARDS

This study was approved by the ethics committees of Leiden University Medical Center.

## 7. ACKNOWLEDGMENTS

We would like to thank the LOLA11 Challenge organizers for the data collection and the creation of ground-truth labels. This work is supported by the China Scholarship Council No.202007720110. The authors report no conflicts of interest.